\documentclass[twocolumn]{openjournal}

\usepackage{lipsum}
\usepackage{amsmath} 

\usepackage{xcolor}
\usepackage{textgreek}
\usepackage[utf8]{inputenc}
\usepackage[english]{babel}
\usepackage{multirow}
\usepackage{hyperref}
\hypersetup{
    unicode, 
    colorlinks=true,
    linkcolor=linkcolor,
    citecolor=linkcolor,
    filecolor=linkcolor,
    urlcolor=linkcolor,
}
\usepackage{color,colortbl}
\definecolor{linkcolor}{rgb}{0.0,0.3,0.5}
\usepackage{tensind}
\tensordelimiter{?}
\DeclareGraphicsExtensions{.bmp,.png,.jpg,.pdf}
\usepackage{verbatim}
\usepackage[normalem]{ulem}
\usepackage{orcidlink}
\usepackage{soul}

\graphicspath{ {./figs/} }

\begin{document}
\title{\vspace{-0.5em}Multiphase Gas in the Ram-pressure-stripped Elliptical Galaxy NGC 1404\vspace{-3em}}
\author{\href{https://orcid.org/0000-0003-0854-8313}{Ryan Eskenasy}$^{*~1}$}

\author{\href{https://orcid.org/0000-0002-3886-1258}{Yuanyuan Su}$^{1}$}

\author{\href{https://orcid.org/0000-0002-4923-396X}{Qinnan Zhu}$^{2}$}

\author{\href{https://orcid.org/0000-0001-5725-0359}{Charles Romero}$^{3,4}$}

\author{\href{https://orcid.org/0000-0001-7557-9713}{Junjie Mao}$^{2}$}

\author{\href{https://orcid.org/0000-0001-5262-6150}{Yuan Li}$^{5}$}

\author{\href{https://orcid.org/0000-0001-8390-9356}{Tirso Marin-Gilabert}$^{6,7}$}

\author{\href{https://orcid.org/0000-0003-3471-7459}{Romana Grossov\'{a}}$^{8}$}

\author{\href{https://orcid.org/0000-0001-6638-4324}{Valeria Olivares}$^{9,10}$}

\author{\href{https://orcid.org/0000-0001-5880-0703}{Ming Sun}$^{11}$}


\author{\href{https://orcid.org/0000-0003-2076-6065}{Elke Roediger}$^{12}$}

\author{\href{https://orcid.org/0000-0002-0765-0511}{Ralph P. Kraft}$^{7}$}

\author{\href{https://orcid.org/0000-0001-8178-8343}{Sheng-Chieh Lin}$^{13}$}
\author{Aaron Burke$^{1}$}


\vspace{1 cm}

\email{$^*$E-mail: reskenasy@uky.edu}

\affiliation{$^{1}$ Department of Physics and Astronomy, University of Kentucky, 505 Rose Street, Lexington, KY 40506, USA}

\affiliation{$^{2}$ Department of Astronomy, Tsinghua University, Beijing 10084, China}

\affiliation{$^{3}$ Department of Physics and Astronomy, University of Pennsylvania, 209 S. 33rd St., Philadelphia, PA 19014, USA}

\affiliation{$^{4}$ ADNET Systems, Inc., 6720B Rockledge Drive, Bethesda, MD 20817, USA}

\affiliation{$^{5}$ Department of Astronomy, University of Massachusetts Amherst, Amherst, MA 01003, USA}

\affiliation{$^{6}$ Universit\"ats-Sternwarte, Fakult\"at f\"ur Physik, Ludwig-Maximilians-Universit\"at M\"unchen, Scheinerstr. 1, 81679 M\"unchen, Germany}

\affiliation{$^{7}$ Center for Astrophysics $|$ Harvard \& Smithsonian, 60 Garden St., Cambridge, MA 02138, USA}

\affiliation{$^8$ Institut de Radioastronomie Millimétrique (IRAM), 300 Rue de la Piscine, 38400 Saint-Martin-d’Hères, France}

\affiliation{$^9$ Departamento de F\'isica, Universidad de Santiago de Chile, Av. Victor Jara 3659, Santiago 9170124, Chile}

\affiliation{$^{10}$ Center for Interdisciplinary Research in Astrophysics and Space Exploration (CIRAS), Universidad de Santiago de Chile, Santiago 9170124, Chile}

\affiliation{$^{11}$ Department of Physics and Astronomy, University of Alabama in Huntsville, Huntsville, AL 35899, USA}

\affiliation{$^{12}$ E. A. Milne Centre, University of Hull, Cottingham Road, HU6 7RX, Hull, UK}

\affiliation{$^{13}$ Departamento de F\'isica Te\'orica, Universidad Aut\'onoma de Madrid, Cantoblanco 28049, Madrid, Spain}

\begin{abstract}
{Galaxy groups and low-mass galaxy clusters are major sites of environment-driven galaxy evolution. NGC~1404 is an X-ray-bright early-type galaxy traveling through the intracluster medium of the nearby low-mass Fornax Cluster, and it hosts off-center warm ionized gas. With minimal ongoing stellar feedback and negligible active galactic nucleus feedback, NGC~1404 provides an ideal laboratory for constraining the impact of environmental processes on the thermodynamics of the interstellar medium. We analyzed VLT/MUSE, 506\,ks {\it Chandra} ACIS, and 119\,ks {\it XMM-Newton} RGS observations to quantify gas motions in both the hot and warm phases of NGC~1404. We probed the kinematics of the warm gas by deriving the velocity structure function (VSF) from MUSE H$\alpha$ integral-field spectroscopy. To measure the kinematics of the hot gas, we derived the power spectrum of X-ray surface brightness fluctuations from the {\it Chandra} data, and independently constrained galaxy-scale hot gas motions by measuring Fe~{\sc xvii} resonant scattering and turbulent line broadening with RGS. Our RGS spectral analysis gives a three-dimensional Mach number of $0.48^{+0.22}_{-0.14}$, consistent with the power-spectrum results from {\it Chandra} for the same extraction region, and corresponding to a turbulent-to-thermal energy density ratio of $\sim0.13$. The warm-gas VSF follows the expectation for Kolmogorov turbulence and is broadly consistent with the hot-gas kinematics. The off-center warm gas in NGC~1404 may have formed through radiative cooling of the hot interstellar medium, triggered by the uplift of low-entropy hot gas during ram pressure stripping. Our study suggests that ram pressure stripping can drive substantial gas motions and reshape the thermodynamic state of the interstellar medium of cluster member galaxies.\vspace{0.5em}}
\end{abstract}


\maketitle

\section{Introduction}
\label{sec:intro}

A galaxy's environment has long been recognized as a key factor in driving its evolution \citep{blanton2009physical}, particularly for those residing in groups and clusters \citep{Su2014,Jain2025}.  
In addition to inhabiting regions of enhanced galaxy density, member galaxies move through the intracluster or intragroup medium (ICM/IGrM) -- a diffuse, X-ray-emitting, ionized plasma. One of the most prominent environmental processes is ram pressure stripping (RPS), in which the drag exerted by the ambient ICM removes gas from a galaxy as it moves through it \citep{gunn1972infall}, effectively displacing its interstellar medium (ISM). Observationally, this process is often characterized by a stripped tail extending downstream and, particularly for early-type galaxies (ETGs), a cold front that marks the boundary between the galaxy’s cooler, denser ISM and the surrounding hotter ICM. 
Such features have been widely used to probe the microphysics of the ICM/IGrM over a broad dynamical range, including viscosity and magnetic fields, which are important for using galaxy clusters as cosmological probes \citep{Su2020,Lin2022,Lin2025}. Meanwhile, infall and stripping processes can significantly modify the galaxy properties, including driving turbulence in the ISM of member galaxies. 

A large fraction of the galaxies found in groups and clusters are ETGs, whose ISM is dominated by hot, X-ray-emitting gas. 
In addition, some of them harbor substantial multiphase gas reservoirs including warm ionized \citep{heckman1989dynamical,hatch2007ionized,werner2014origin, tremblay2018galaxy,Olivares_2019} and molecular gas \citep{edge2001detection,salome2003cold,lim2008radially}. The origin and kinematics of the multiphase gas in ETGs have been studied extensively in the context of mechanical feedback from central supermassive black holes (SMBHs), particularly in brightest cluster galaxies (BCGs) \citep{fabian2003relationship,li2014modeling,russell2019driving,Li2020,Olivares_2019}. One origin of the cooler gas may be radiative cooling and condensation of the hot atmosphere, potentially through thermal instabilities \citep{field1965thermal,balbus1986local}. In a stably stratified atmosphere, buoyant motions can oppose the nonlinear development of condensates, motivating comparisons between the cooling and dynamical timescales of the gas to probe the conditions favorable for multiphase condensation \citep{mccourt2012thermal,sharma2012thermal,voit2017global}.

In observations, H$\alpha$ filaments are often found to be dragged up underneath X-ray cavities, motivating an interpretation where rising AGN-driven bubbles uplift low-entropy gas to larger radii where it may cool and condense \citep{mcnamara2016mechanism}. Turbulence driven by outbursts from SMBHs has also been considered a promising channel for transporting mechanical energy to heat the ambient hot gas.


{\it XRISM}-Resolve is now transforming this field by providing direct measurements of hot gas kinematics, thanks to its high spectral resolution. Subsonic, turbulent hot gas motions with line-of-sight velocity dispersions $\sim100-200$ km s$^{-1}$ were found at the centers of relaxed cool core clusters, 
corresponding to modest turbulence pressure relative to the thermal pressure \citep{xrism2025xrism, xrism2025bulk, rose2025xrism, xrism2026disentangling}. 
However, these measurements are largely limited to cluster centers and cannot yet probe the kinematics of the ISM of non-BCG galaxies, ICM at cluster outskirts, or hot gas at small scales due to the modest spatial resolution of {\it XRISM}-Resolve and its limited sensitivity to soft X-rays. This motivates alternative approaches, including deriving the power spectrum from surface-brightness fluctuations \citep{churazov2012x,Romero_etal_2024}, utilizing the dependence of resonant scattering on turbulence \citep{2017MNRAS.472.1659O}, and using the kinematics of warm gas as a tracer of that of hot gas \citep{Li2020,2023FrASS..1038613G}.

\citet{Eskenasy_2023,Eskenasy_2026} studied the warm nebulae in ETGs that are not at the centers of clusters or groups, based on Integral Field Unit (IFU) observations obtained with the Multi Unit Spectroscopic Explorer (MUSE) on the Very Large Telescope (VLT). VLT/MUSE features broad spectral coverage with superb spectral and spatial resolution and a large field of view, enabling detailed and robust studies of warm ionized gas, which are not achievable with previous IFU surveys, e.g., SAURON \citep{Bacon_2001}, CALIFA \citep{sanchez2012califa,walcher2014califa}, SAMI \citep{croom2012sydney,bryant2015sami}, and MaNGA \citep{bundy2014overview}. 
Unlike the warm gas in BCGs, which is predominantly filamentary, warm gas in non-BCG ETGs is found to display a more diverse morphology \citep{Eskenasy_2026}. 
Some galaxies display rotating disks that are morphologically and kinematically aligned with their stellar components, resembling late-type galaxies. 
Others exhibit filamentary nebulae that are kinematically decoupled from the stellar body, resembling the structures observed in the BCGs of cool-core clusters \citep{Olivares_2019}. 
A few filamentary systems, however, exhibit off-center warm nebulae, with no emission-line gas detected at the galaxy centers. The diverse morphology of warm nebulae in non-BCG ETGs may reflect multiple channels for multiphase gas formation in them.  

Compared to active galactic nuclei (AGN)-driven processes, environment-driven ISM dynamics and multiphase gas formation remain relatively understudied. A key challenge is disentangling external effects from internal processes, as both typically coexist in most galaxies. 
NGC~1404, one of the nearest ($D=18.8$ Mpc, 1$^{\prime\prime}=0.092 ~\mathrm{kpc}$) X-ray bright elliptical galaxies, may provide the best opportunity to probe the environmental effects.  
It is currently moving through the ICM of the Fornax Cluster, which is a low-mass cluster with an ICM temperature of 1.5\,keV, representing the environments of most nearby galaxies \citep{Su2017Fornax}. 
Deep X-ray observations have revealed 
clear signs of ram pressure stripping including a sharp cold front at the leading edge and a tail in the downstream region \citep{machacek2005infall,Su_etal_2017,Su2017}. Furthermore, simulations tailored to the specific infall scenario of NGC~1404 through Fornax demonstrate that the flow around NGC~1404 is heavily perturbed and its downstream wake should exhibit enhanced turbulent gas motions \citep{sheardown2018recent}. To the best of our knowledge, NGC~1404 is not known to host clear signatures of AGN feedback such as X-ray cavities and radio jets \citep{grossova2022very}, making it a clean laboratory for isolating how the environment can drive and regulate the kinematic structure in the hot phase. This system is additionally compelling in that it is one of the few systems displaying off-center warm-ionized gas identified by \cite{Eskenasy_2026}. 
In principle, such warm gas may have been brought in externally via mergers with gas-rich systems, or formed through condensation of the galaxy's hot ISM -- either locally or following the displacement of low-entropy central gas by RPS.

In this paper, we combine IFU observations of NGC~1404 obtained with MUSE with deep X-ray datasets from \emph{Chandra} and \emph{XMM-Newton}. Our goal is to probe the role of ram pressure stripping in shaping 
the multiphase gas in ETGs. 
Previously, the hot gas kinematics of NGC~1404 were studied using the Reflection Grating Spectrometer (RGS) on board {\it XMM-Newton} by measuring line broadening and line ratios sensitive to resonant scattering \citep{2017MNRAS.472.1659O}, which found an exceptionally turbulent hot ISM. 
The X-ray observation we use for this study is more than five times deeper. 
In Sec.~\ref{sec:data}, we describe the MUSE, \emph{Chandra}, and \emph{XMM-Newton} datasets and reduction. The warm gas analysis as well as the hot gas velocity analysis is given in Sec.~\ref{sec:analysis}. We discuss the physical implications of the hot-warm gas analysis in Sec.~\ref{sec:discussion} and provide concluding remarks in Sec.~\ref{sec:conclusions}.


\section{Observations and Data Reductions}
\label{sec:data}

The observational logs of VLT/MUSE, {\it Chandra}/ACIS, and {\it XMM-Newton}/RGS are listed in Table~\ref{tab: xmm-record}. Below we describe the details of the data reduction. 

\subsection{VLT/MUSE} 

Archival VLT/MUSE integral-field spectroscopic observations of NGC~1404 were used to characterize its nebular emission and stellar properties. MUSE is a wide field of view (1$^{\prime}~\times~$1$^{\prime}$), image-slicing IFU spectrograph with broad optical wavelength coverage. The data were reduced using the ESO MUSE pipeline via the European Southern Observatory Recipe Execution Tool (ESOREX v.3.13.1; \citealp{weilbacher2020data}). This includes bias subtraction, flat-fielding, wavelength/flux calibration, and reconstruction of a final datacube with two spatial axes and a third wavelength axis. The final datacube was sky subtracted using version 2.1 of the Zurich Atmosphere Purge package (ZAP v2.1; \citealp{soto2016zap}) in addition to the data reduction pipeline's internal sky subtraction method. The datacube is corrected for galactic foreground extinction using the dust maps of \cite{schlegel1998maps}. The final reduced MUSE cube spans $\sim4750-9300 \text{\AA}$.

Higher-level data products were generated by fitting the stellar continuum and nebular emission at each spaxel. The stellar redshift was estimated using template matching and cross-correlation with the \verb|AUTOZ| \citep{baldry2014galaxy} code. Stellar velocity dispersions were then obtained using \verb|VDISPFIT|. We then used the \verb|PLATEFIT| spectral fitting code \citep{brinchmann2004physical,tremonti2004origin} to independently model the stellar continuum and nebular emission. After masking potential emission lines, a series of stellar population templates and the previously derived stellar kinematics were used to model the underlying stellar continuum. The best-fit continuum was then subtracted off and the residual spectrum was used to fit the nebular emission lines. Potential emission lines were modeled with a single Gaussian profile and were all tied to share common velocities and velocity dispersions.

\begin{figure*}[t]
    \centering
    \includegraphics[width=1.0\textwidth]{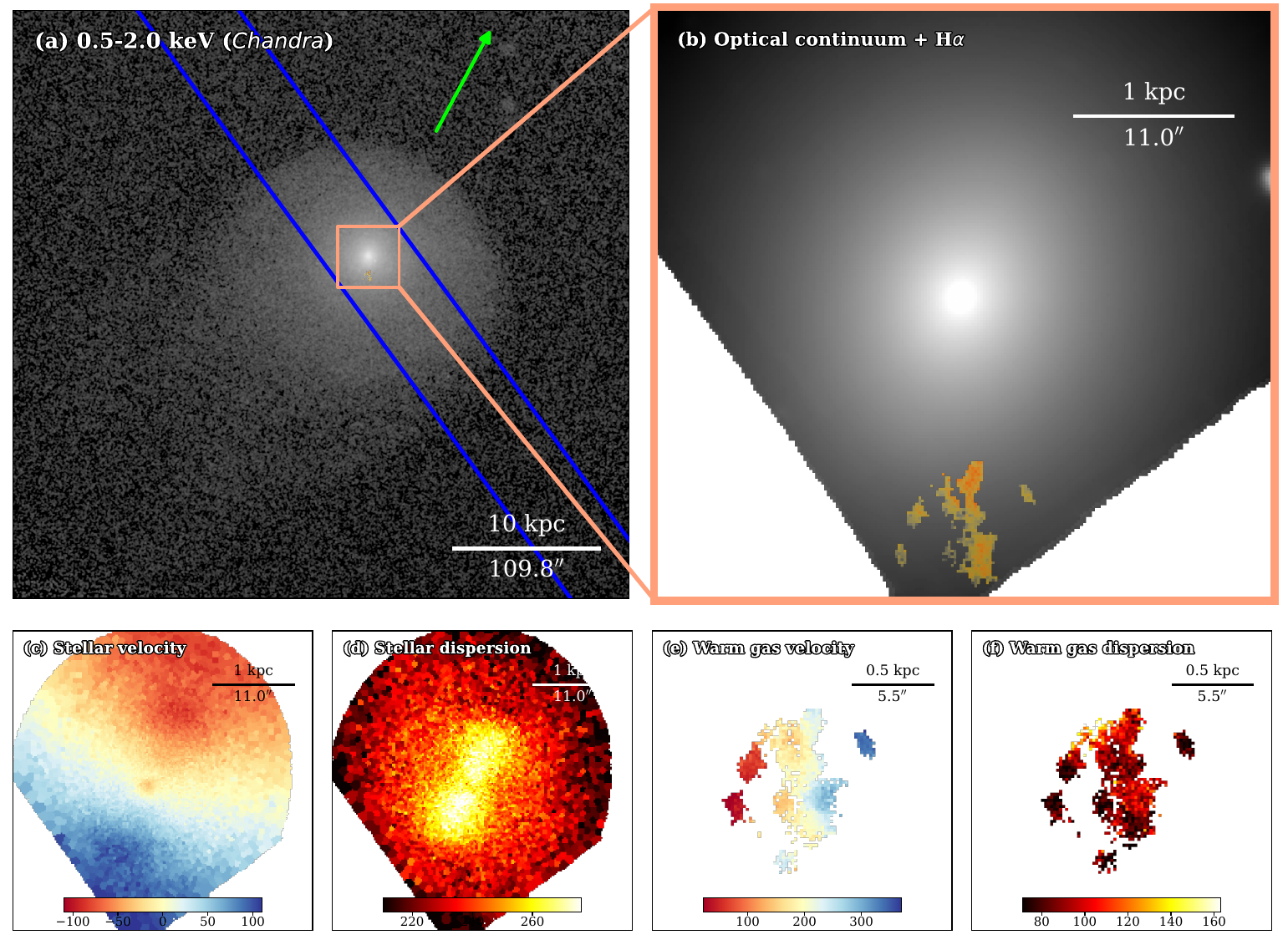}
    \caption{
    Panel (a) \emph{Chandra} image of NGC~1404, overlaid with H$\alpha$ emission in yellow and orange. The green arrow is in the direction of NGC~1399. The blue lines trace the RGS extraction aperture The pink lines trace the field of view captured in panel (b). Panel (b) shows the MUSE optical continuum image centered on NGC~1404, overlaid with the H$\alpha$ emission in yellow and orange. Panels (c) and (d) show the stellar line-of-sight velocity and velocity dispersion, respectively,
    Panels (e) and (f) show the warm gas line of sight velocity and velocity dispersion, respectively, both centered and zoomed in the warm nebula. The physical scale is indicated in each panel.}
    \label{fig:continnum}
\end{figure*}

To account for the spurious detections within the emission line maps, we only include spaxels where the emission line signal-to-noise ratio $>7$ and the inferred gas velocity dispersion $>40$ km s$^{-1}$. Additionally, we filtered the data by excluding spaxels where the 1$\sigma$ uncertainty on the gas velocity and velocity dispersion exceeds a given threshold. These cut-offs were determined by calculating the average velocity and velocity dispersion errors in an offset, emission-free region in the field. MUSE optical continuum and H$\alpha$ images of NGC 1404 are shown in Fig.\ref{fig:continnum}.


\subsection{Chandra} 
We included 506 ksec {\it Chandra} ACIS observations, which were selected to be focused on the center of NGC~1404 with no CCD gaps within regions of interest. 
We used \texttt{CIAO 4.17} and \texttt{CALDB 4.12.0} to process and reduce the {\it Chandra} data. Level 1 events were reprocessed using \texttt{chandra\_repro}. Light curve flares beyond $3\sigma$ were filtered using \texttt{lc\_clean}. Point sources were detected in a broad band image with \texttt{wavdetect} at scales of 1,2, and 4 pixels. We extracted an image in the 0.5--2.0\,keV energy band, merged from all observations. We replaced point sources with pixel values interpolated from surrounding background
regions using \texttt{dmfilth}. 
The combined instrumental and astrophysical background was approximated by the blank-sky fields, normalized by the count rate in the 9.5--12.0\,keV energy band relative to each observation and image using \texttt{blanksky} and \texttt{blanksky\_image}. A point source removed and background-subtracted image of NGC~1404 is shown Fig.~\ref{fig:continnum}, panel (a).

\begin{figure}
\centering
   \includegraphics[width=0.95\linewidth]{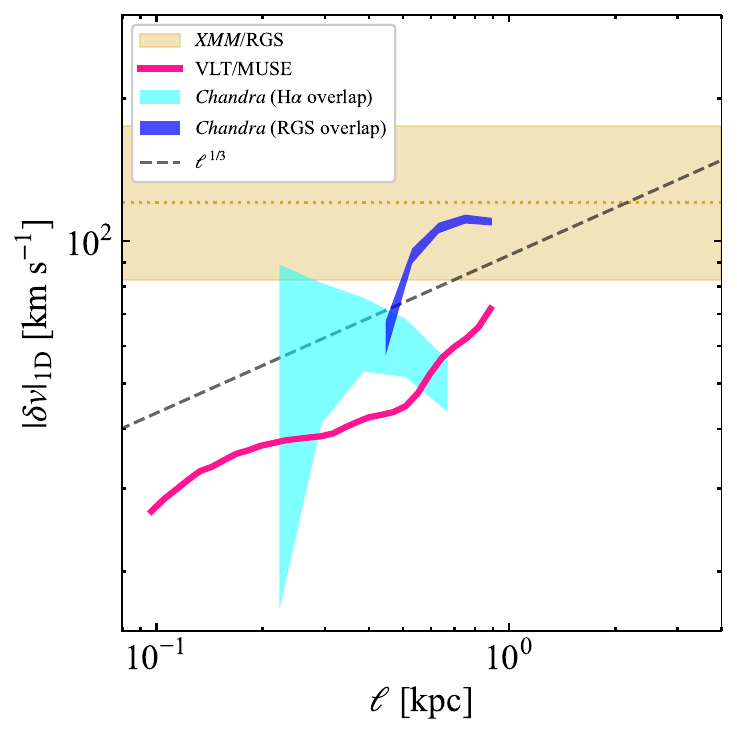}

   \caption{Comparison of the warm and hot gas motions as a function of spatial scale. The velocity structure function of the warm nebula derived from MUSE observations (rotation removed) is shown with the magenta curve. The cyan band shows the velocity amplitudes of the hot gas inferred from the \emph{Chandra} surface brightness fluctuations in the region directly containing the warm nebula. 
   The uncertainties shown are due to Poisson noise only. 
   The blue curve traces the corresponding \emph{Chandra} velocity amplitudes within the rectangular RGS extraction region.  
   The gold dotted horizontal line and shaded band mark the median and central 68$\%$ confidence interval of the one-dimensional RGS velocity constraint estimated from the joint resonant scattering and line broadening analysis. The dashed grey line denotes the Kolmogorov classical scaling $|\delta v|\propto\ell^{1/3}$ (with arbitrary normalization).}
    \label{fig:vel}
\end{figure}


\begin{figure*}
    \centering   \includegraphics[width=0.95\linewidth]{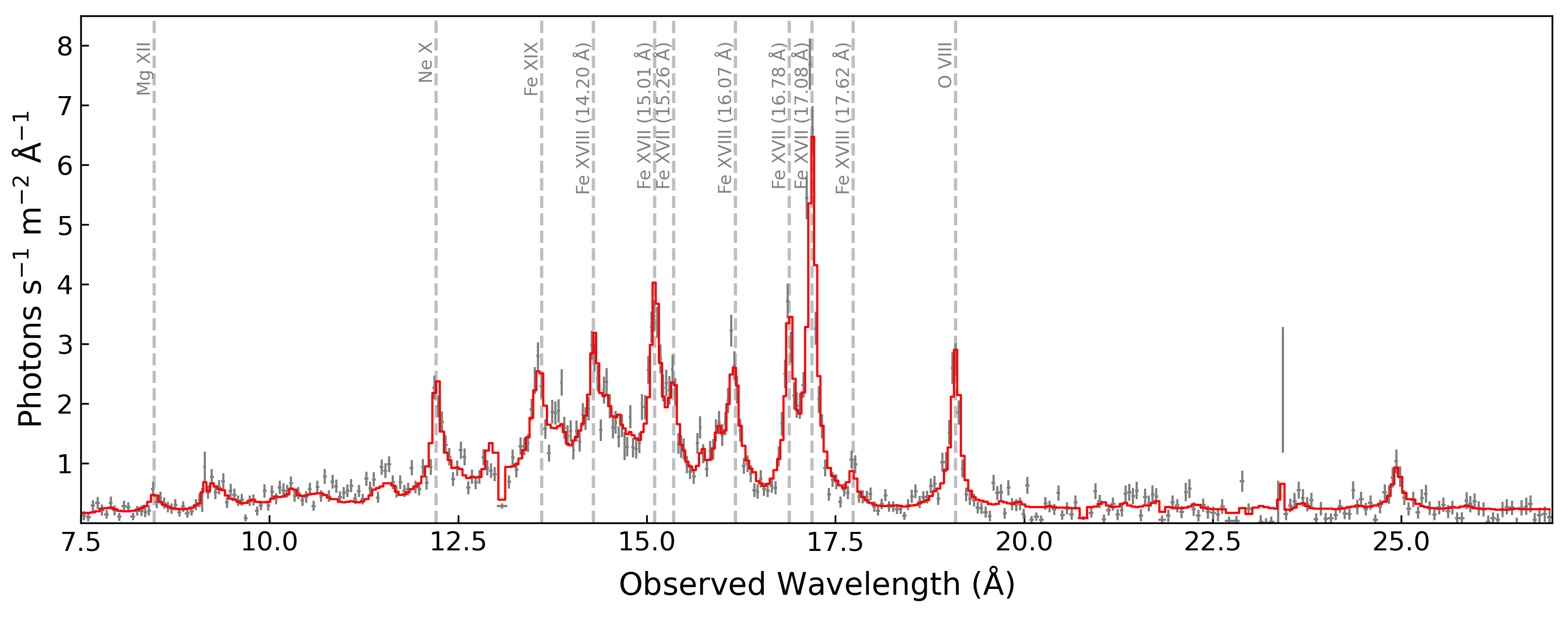}
    \caption{RGS spectra of observation 0781350101 extracted from 90\% PSF region overplotted with the best-fit model with $s$ frozen to 1.}
    \label{fig: frgs-spectral-fitting}
\end{figure*}

\subsection{XMM-Newton/RGS} 
We reduced two archived RGS observations of NGC 1404 obtained by {\it XMM-Newton}. The RGS spectra were extracted with XMM-Newton Science Analysis System (SAS) v22.1.0 following \citet{mao2023xmmnewtonreflectiongratingspectrometer}. To account for contamination by soft protons, we extracted the light curve from CCD 9, where the fewest source photons were recorded. The good time intervals (GTIs) were then selected with the \verb|deflare| task in CIAO using sigma-clipping. Since observation 0304940101, presented in \citet{2017MNRAS.472.1659O}, was significantly affected by the soft-proton flare, we reassigned the initial value in the \verb|deflare| task to obtain an acceptable selection. However, some of the GTIs of Observation 0304940101 still overlapped with flare intervals. The spectra were extracted from a 90\% PSF region ($\sim1^{\prime}$, following \citealp{2022MNRAS.511.3159M}).
When compared with the new observation 0781350101, spectra extracted from observation 0304940101 display an elevated normalization, which may be partially attributed to the residual flares. Because the effective exposure time of observation 0781350101 is more than five times deeper than Observation 0304940101, our analysis is restricted to observation 0781350101.

\section{Analysis and Results}
\label{sec:analysis}
\subsection{Emission line analysis and warm gas properties}

\subsubsection{Velocity Structure Function}
NGC~1404 hosts a prominent warm ionized nebula, offset by $\sim1.5\,\mathrm{kpc}$ to the south of the center of the galaxy (Fig.~\ref{fig:continnum}, panel b). Fainter gas clouds may be present north to the center of the galaxy, though these features are not securely recovered using our conservative emission-line selection criteria. The line of sight velocity field of the warm gas (Fig.~\ref{fig:continnum}, panel e) exhibits ordered rotation along the east-west direction. Its rotation is misaligned with that of the stars (Fig.~\ref{fig:continnum}, panel c).

To probe the level of turbulence of the warm ionized gas, we compute the first-order velocity structure function (VSF) of the line-of-sight velocity map of emission lines (with their velocities linked in the spectral fitting), following \cite{Li2020}. For each pair of pixels, we obtain the spatial separation in the plane of the sky, $\ell$, and their velocity difference, $\delta v$. The VSF is then derived from the average absolute value of the velocity differences, $\langle\delta v\rangle$, at each scale of $\ell$. The warm nebula's velocity gradient likely dominates the VSF over turbulent contributions. We modeled the gradient plane in the velocity map using $a x+ by+c$, which was then subtracted from the velocity map. 
The resulting VSF of the warm nebula is shown in Fig.~\ref{fig:vel} (pink). 
The smallest separation plotted is set by the seeing limit. At large separations, the VSF is averaged over a small fraction of the entire volume. Bins with numbers of pairs below 20\% of the peak have large sampling uncertainties and are not plotted. Over the range of separation we probe, its VSF agrees well with the expectation for classical Kolmogorov turbulence. 

\subsubsection{Emission-line diagnostics}

The broad wavelength range of our MUSE spectra can be used to measure several key optical line ratios that are sensitive to various ionization mechanisms. 
We utilize the Baldwin, Phillips, and Terlevich (BPT) diagram \citep{baldwin1981classification} to compare the [NII]/H$\alpha$ and [OIII]/H$\beta$ line ratios of the warm gas to distinguish the dominant source of ionization \footnote{spaxels are included only if all four emission lines have SNR $>7$}.

\begin{figure}
\centering
\includegraphics[width=1\linewidth]{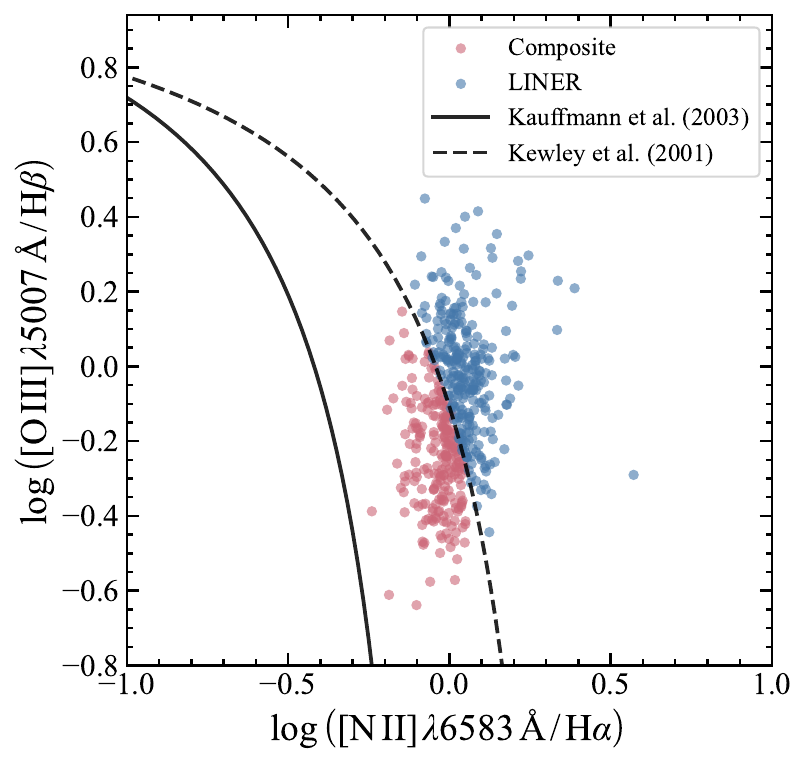}
\caption{BPT diagram for spaxels in NGC~1404's MUSE datacube. The dashed black curve is the theoretical upper boundary for starbursts from \cite{kewley2001theoretical} and the solid black curve is the empirical demarcation from \cite{kauffmann2003host}. The red points falling in between the curves occupy the composite region. The blue points lie in the LINER-like region. The bulk of the spaxels are clustered close to the \cite{kewley2001theoretical} demarcation curve, spanning the composite and LINER-like zones.}
\label{fig:bpt}%
\end{figure}

Fig.~\ref{fig:bpt} shows NGC~1404's BPT diagram. The spaxel distribution is preferentially clustered between the standard composite and Low-Ionization Nuclear Emission-Line Region (LINER) zones, with many spaxels branching farther into the LINER zone. This indicates that the warm nebula in NGC~1404 is unlikely to be powered by a classical HII-region radiation field. 
We adopt the LINER metallicity calibration in the following section as a first-order description of the seemingly low-ionization conditions, as suggested by the BPT diagram. 


\subsubsection{Metallicity}

The gas-phase oxygen abundance was calculated using emission line ratios. 
For the LINER-like spaxels, the calibration presented in \cite{kumari2019metallicity} was used:

\begin{equation}
\begin{aligned}\label{eq:(2)}
12 + {\rm log(O/H)} &= 7.673
+ 0.22 \cdot \sqrt{25.25 - 9.073~ (\mathcal{O}-\mathcal{N})}\\
&\quad + 0.127~\mathcal{O}\\
\mathcal{O} &\equiv {\rm log}\left(\frac{{\rm [OIII]}\lambda5007}{{\rm H}\beta}\right)\\
\mathcal{N} &\equiv {\rm log}\left(\frac{{\rm [NII]}\lambda6583}{{\rm H}\alpha}\right)
\end{aligned}
\end{equation}





The 16\%--84\% percentile distribution of the oxygen abundances of all spaxels is 8.77 to 8.81 with a median of 8.79 as shown in Table~\ref{tab:phase_properties}. 

The radial Fe metallicity profile of the hot gas in NGC~1404 was studied in detail by \cite{2022MNRAS.511.3159M}, accounting for the multi-temperature structure of its hot atmosphere. We adopt their Fe metallicity of $Z_{\mathrm{Fe}} = 0.79\pm0.08$ at the radius where the warm gas resides. \cite{2022MNRAS.511.3159M} measured an oxygen-to-iron ratio of $0.74 \pm 0.06$, whereas in this work we find a ratio of $\sim1.17$ based on our RGS spectral analysis. Combining our oxygen-to-iron ratio with their Fe metallicity, we find $12+\log(\mathrm{O/H})=8.75\pm0.06$ for the hot gas. 
Accounting for the systematic difference between our oxygen-to-iron ratio and that of \cite{2022MNRAS.511.3159M}, we find an inferred range for the oxygen abundance of $\sim8.5-8.79$, as shown in Table~\ref{tab:phase_properties}.
The oxygen abundance of the warm ionized gas is comparable to the hot phase at the same radius from the galaxy center, if adopting the upper end of the hot gas abundance range. However, we find that the warm gas could be enriched relative to the hot phase by up to a factor of $\sim2$. 


\subsubsection{Density and Pressure}

The electron density ($n_e$) can be traced by the [S II]$\lambda$6716/[S II]$\lambda$6731 flux ratio as increasing electron density enhances collisional de-excitation, preferentially suppressing the line with the lower critical density ([S II]$\lambda$6716), thereby altering the line ratio. 
The $n_e$ of NGC~1404's warm gas was determined from each spaxel's [S II]$\lambda$6716/[S II]$\lambda$6731 flux ratio using PyNeb \citep{luridiana2015pyneb}. We assume a uniform electron temperature $T_e=10^4$ K, though slight temperature variations are likely and should be considered when interpreting the derived $n_e$. Nevertheless, this temperature is appropriate for warm, photoionized nebulae. As the [SII] flux ratio becomes insensitive to $n_e$ near the low-$n_e$ and high-$n_e$ limits, we adopt the filtering procedure used in \cite{jiang2019correlation}. Using their method, we include spaxels only if their [SII] flux ratio (including its uncertainty interval) falls within the $n_e$-sensitive regime. Spaxels that are consistent with either of the asymptotic limits are excluded from our set of secure $n_e$ estimates. Using 151 spaxels with secure estimates, we find a median $n_e = 842^{+1046}_{-371} ~\mathrm{cm^{-3}}$ (16th-84th percentiles). To estimate the characteristic thermal pressure ($P$) of the fully ionized warm gas, we assume a constant $T_e\sim10^4 $ K, and find $P\sim2\times10^{-9}$ erg cm$^{-3}$.

\subsection{Hot Gas Surface Brightness Fluctuation}
The amplitude of the power spectrum of X-ray surface-brightness fluctuations in the hot gas is expected to be linearly proportional to the level of turbulent motion \citep{Churazov_etal_2012, Arevalo_etal_2012}. 
We extract the power spectrum of the X-ray surface brightness
fluctuations of the ISM of NGC~1404, following \citep{Romero_etal_2023, Romero_etal_2024, Romero_etal_2025, Eckert_etal_2017}. 

We obtain a best-fit $\beta$ profile, $I_0  (1 + (r / r_c)^2)^{0.5 - 3\beta}$, with $\beta=0.48$ and $r_c=6.1$\,pixels for the surface-brightness distribution over a radial range of 5–150 pixels. We excluded the central region ($r < 5$\,pixels) from the analysis, as its surface-brightness profile deviates from a single $\beta$ model.
A residual image, $I_{\rm res}$, is constructed by subtracting from the background-subtracted source image, $I_{\rm src}$, a $\beta$-model image multiplied by the exposure map as $I_{\rm res}=I_{\rm src} - I_{\rm beta} \times E$. 
We then obtain a convolved image from the residual image, $I_{\rm res}$, and the exposure map, $E$,

\begin{equation}
    I_{\sigma}=\left(\frac{G_{\sigma_1} \circ I_{\rm res}}{G_{\sigma_1} \circ E}-\frac{G_{\sigma_2} \circ I_{\rm res}}{G_{\sigma_2} \circ E}\right)\times E
\end{equation}
where $G_{\sigma_1}$ and $G_{\sigma_2}$ are Gaussian filters at widths of $\sigma_1={\sigma}/\sqrt{1+\epsilon}$ and $\sigma_2={\sigma}\sqrt{1+\epsilon}$, with $\epsilon\ll1$, and $\sigma$ and the wavenumber, $k$, are related by $\sigma=1/\sqrt{2\pi^2k^2}$. We then calculate the variance of $I_{\sigma}$ for each $k$ as $V_k=\sum{{I_\sigma}^2}$ and the mean squared $\beta$-model surface brightness, $\overline{{\mu_{\rm S}}^2}$. The 2D power spectrum can be derived as follows:
\begin{equation}
P_{\rm 2D} (k)=\frac{1}{\epsilon^2\pi k^2\overline{{\mu_{\rm S}}^2}}\frac{V_k}{\sum{E^2}}.
\end{equation}

We calculate the associated Poisson noise as 
\begin{equation}    
P_{\rm poisson}=\frac{1}{{\mu_S}^2}\frac{\sum I_{\rm src}}{\sum{E^2}}.
\end{equation} 
The Poisson noise-subtracted projected 2D power spectra of surface brightness fluctuations are related to the 3D power spectra of X-ray emissivity fluctuations through a window function
\begin{equation}
P_{\rm 2D} (k_r) =\int P_{\rm 3D} ({\rm\mathbf{k}}) (|\overset{\sim}{W}(k_z)|^2) dk_z,
\end{equation}
where $z$ is the axis along the line-of-sight, $r^2=x^2+y^2$ is in the plane of the sky, and $|\overset{\sim}{W}(k_z)|^2$ is the one-dimensional power spectrum of the window function, which normalizes the contribution of the relevant 3D signal (emissivity) fluctuations to the 2D surface brightness. As demonstrated in \cite{Romero_etal_2023}, the above equation can be approximated by 
\begin{equation}
P_{\rm 2D} (k_r) \approx N P_{\rm 3D} ({\rm\mathbf{k}}), 
\end{equation}
where $N=\int |\overset{\sim}{W}(k_z)|^2 d k_z$.

We calculate $\overset{\sim}{W}(k_z)$ from the Fourier transform of the line-of-sight emission, normalized by the $\beta$-model surface brightness in each annulus. $\overset{\sim}{W}(k_z)$ is integrated along the line-of-sight to obtain $N(r)$ for each narrow radial bin. We then obtain a single $\langle N\rangle$ by taking the area-weighted average of $N(r)$ over all radial bins.

The emissivity 3D power spectra were used to calculate the 3D amplitude of the surface brightness fluctuations as 
$A_{\rm 3D}=\sqrt{P_{3D}(k)4\pi k^3}$. 
The amplitude of the density fluctuation was calculated as $\delta{\rho}/\rho=A_{\rm 3D}/2$, which can be used to probe the one-dimensional Mach number assuming $\mathcal{M}_{\rm 1D}\approx \delta{\rho}/\rho/\eta$. Cosmological simulations have calibrated the proportionality coefficient to be $\eta \approx 1$ for the ICM of relaxed clusters \citep{Zhuravleva_etal_2023}. 
To the best of our knowledge, this coefficient has not yet been calibrated for the hot ISM of early-type galaxies.
In the analysis below, we nevertheless adopt $\eta=1$. 
The ISM of NGC~1404 has a temperature of $\approx0.7$\,keV, for which we derived a sound speed of $c_s=\sqrt{\gamma k_\mathrm{B}T/\mu m_p}=430$\,km/s, where $\gamma=5/3$ is the adiabatic index for a monatomic ideal gas and $\mu=0.62$ is the average molecular weight. We then calculated the one-dimensional velocity for the hot ISM of NGC~1404 using $v_{\rm 1D}=c_s \mathcal{M}_{\rm 1D}$. 

We performed the analysis to probe the hot-gas kinematics 
for a slab intended to overlap the RGS aperture (overlaid in Fig.~\ref{fig:continnum}, panel (a)). As shown in Fig.~\ref{fig:vel}, 
the good agreement between the hot gas velocities measured with RGS and probed with {\it Chandra} supports our adoption of $\eta \approx 1$. We further performed the above analysis for eight circular regions of equal aperture radius and projected distance to the center of NGC~1404. 
One of these circular regions directly overlaps with the warm ionized gas, while the other seven are at successive $45^{\circ}$
 offset azimuthal positions around the center of the galaxy. The region overlapping with the warm gas and the adjacent region to the east show characteristic one-dimensional hot-gas velocity amplitudes 
 of $\gtrsim50\,\mathrm{km\,s^{-1}}$, while the other circular regions have velocity amplitudes of $\approx20\,\mathrm{km\,s^{-1}}$. However, given the large uncertainties of the individual measurements, this difference is not statistically significant. 

\subsection{RGS Measured Hot Gas Kinematics} 

\subsubsection{Spectral fitting and line broadening}
Spectral fitting was performed in SPEX (SPEXACT v3.08.01) using the proto-solar-abundance table of \citet{2009LanB...4B..712L}. For extended sources, emission lines of RGS spectra are broadened by $\Delta\lambda=0.138\Delta\theta~\rm\AA$ \citep{2004A&A...420..135T}. To account for this effect, we use a $0.3-2~{\rm keV}$ MOS1 image binned by 80 ($4''$ each bin) to generate line broadening profiles. This broadening is accounted for with one $\textit{lpro}$ component.

The ICM emission is modeled with the collisional ionization equilibrium (\textit{cie}) component. The emission measures, temperature, RMS velocity, and the abundances of N, O, Ne, Mg, and Fe are set free. The Ni abundance is coupled to Fe. We remove all lines associated with the Fe {\sc xvii} ion from the \textit{cie} model due to the unaccounted resonant scattering. Instead, we add 4 delta lines fixed at $15.01~\rm\AA$, $15.26~\rm\AA$, $16.78~\rm\AA$, and $17.08~\rm\AA$, representing the main Fe {\sc xvii} transition lines in this band, where the impact of resonant scattering is the greatest. These four delta lines are broadened by a Gaussian component, whose velocity is coupled to $v_{\rm rms}$ in the \textit{cie} model. These source models are then corrected for redshift, Galactic absorption, and line broadening (\textit{lpro}). The best-fit results are shown in Table~\ref{tab: rgs fitting result-PSF90}. 

Since the spatial broadening and turbulence broadening produce similar effects on the observed line widths, their contributions are partially degenerate in the fit. As a result, the $s$ parameter of \textit{lpro} is strongly correlated with $v_{\rm rms}$.  
Small changes in $s$ can lead to large differences in the inferred $v_{\rm rms}$. If the scale factor is allowed to vary freely, the fitted $v_{\rm rms}$ reaches $336^{+113}_{-155}~{\rm km/s}$ while $s=0.78^{+0.10}_{-0.09}$. However, when the scale factor is fixed to 1, the best-fit $v_{\rm rms}$ becomes much smaller. We evaluate the statistical preference using the Akaike Information Criterion (AIC, \citealt{burnham2002model}). For C-statistics, the AIC is computed as $\rm AIC=2k+C$, where k is the number of free parameters, and C is the minimum value of the fit statistic. The AIC difference, $\Delta_i={\rm AIC}_i-{\rm AIC_{min}}$, is widely used for model comparison. When $\Delta_i$ is smaller than 2, the two models are thought to be indistinguishable; when $\Delta_i$ is larger than 10, the one with the larger AIC should be rejected. The $\Delta{\rm AIC}$ between the model with $s$ fixed to 1 and that with $s$ free is 3, indicating that the two models are close to being statistically indistinguishable. Although the wavelength broadening caused by turbulence, $\Delta\lambda_{v}$, varies with wavelength, while the spatial broadening, $\Delta\lambda_{s}$, is wavelength-independent, the spectral quality is insufficient to robustly separate the contributions of these two components.

The variation in scale factor originates from differences in the broadening profiles derived from different energy bands. 
We generate the line broadening profiles from $0.6-0.7~{\rm keV},~0.7-0.8~{\rm keV}$, and ${0.8-0.9~{\rm keV}}$ narrow band MOS1 images, where the main Fe and O lines are located. By comparing their cumulative distribution function with that of broad-band image, we find corresponding scaling parameters in the range of 0.85 - 1.2. Moreover, in the $0.7-0.8~{\rm keV}$ and $0.8-0.9~{\rm keV}$ bands, where the strongest lines are present, the corresponding scale factors are slightly larger than 1. Therefore, the best-fit value of $s$ is likely underestimated, and fixing $s$ to 1 is more physically motivated. The SPEX \texttt{error} command provides a $1\sigma$ upper limit of $v_\mathrm{rms}<177~\mathrm{km}~\mathrm{s}^{-1}$. 

\subsubsection{Resonant Scattering}

In addition to line broadening, the Fe {\sc xvii} line flux ratio, ${\rm FR}=(I_{17.05\ \rm \AA}+I_{17.10\ \rm \AA})/I_{15.01\ \rm \AA}$ could also be used to constrain turbulence \footnote{The Fe~{\sc xvii} 17.05~\AA\ and 17.10~\AA\ transitions are unresolved in our RGS spectra; their combined flux is represented by the 17.08~\AA\ component in our RGS spectral fitting.} \citep{2017MNRAS.472.1659O}. While the diffuse, hot ISM is generally considered to be optically thin, for an atomic transition with a large oscillator strength such as the 15.01~\AA\ line (the strongest resonance line in the Fe {\sc xvii} complex), the absorption cross section can become large enough such that the hot gas becomes optically thick to that specific photon energy. A second line from the same ion that is produced under the same physical conditions, but with a much smaller oscillator strength, can act as an ``unaffected'' reference. Deviation of the observed line ratio from the intrinsic emissivity ratio, $R_{\rm thin}$, would indicate that resonant scattering suppresses the 15.01~\AA\ line. We expect the observed line ratio to decrease monotonically with the turbulence level of the hot gas, since stronger gas motions broaden the resonance line and reduce suppression. 

We implement a Monte Carlo radiative transfer simulation to predict how the Fe {\sc xvii} 15.01\,\AA\, line is suppressed relative to the sum of the two reference lines with much smaller oscillator strengths (17.05\,\AA and 17.10\,\AA), as a function of the turbulent Mach number in NGC~1404's hot ISM. We obtain Fe {\sc xvii} level and transition data from AtomDB and obtain oscillator strengths of 2.49, 0.126, and 5.2e-8 for the 15.01~\AA, 17.05~\AA, and 17.10\,\AA\, lines, respectively, from the tabulated Einstein A coefficients. 
The 17.10\,\AA\, line is a forbidden line with a very small oscillator strength; thus, its resonant optical depth is negligible, and we treat this line as non-scattering. The $15.01~\AA$ and $17.05~\AA$ lines are treated as lines that scatter, with the $15.01~\AA$ line having the larger oscillator strength of the two. A CIE spectral model for the global temperature of NGC~1404 was used to generate the intrinsic emissivities of all three lines, giving the line ratio that would be observed with zero scattering $R_{\rm thin}=1.47$. We obtain density and temperature profiles from the \emph{Chandra} observations, combined with the ionization balance tables from AtomDB, to obtain the local Fe {\sc xvii} ionic density at any point within the gas. We compute the local line-center cross-section by combining the thermal Doppler broadening (from the local gas temperature) with turbulent broadening via
\begin{equation} \Delta E_D = \frac{E_0}{c} \sqrt{\frac{2 k_B T}{A m_p} + \frac{2}{3} v_{\rm turb}^2} 
\end{equation} 
\begin{equation} \sigma_0 = \frac{\sqrt{\pi}\, h\, c\, r_e\, f_{\rm osc}}{\Delta E_D}
 \end{equation} 
where $A = 55.845$ is the atomic mass number of Fe, $E_0$ the line rest-frame energy, $k_B T$ the local thermal energy, $m_p$ the proton mass, $c$ the speed of light, $v_{\rm turb}$ the turbulent velocity, $h$ Planck's constant, $r_e$ the classical electron radius, and $f_{\rm osc}$ the oscillator strength of the transition. $v_{\rm turb}$ is related to the 3D Mach number by the sound speed, $v_{\rm turb}=\mathcal{M} c_s$. 

\begin{figure*}
   \centering
    \includegraphics[width=0.495\linewidth]{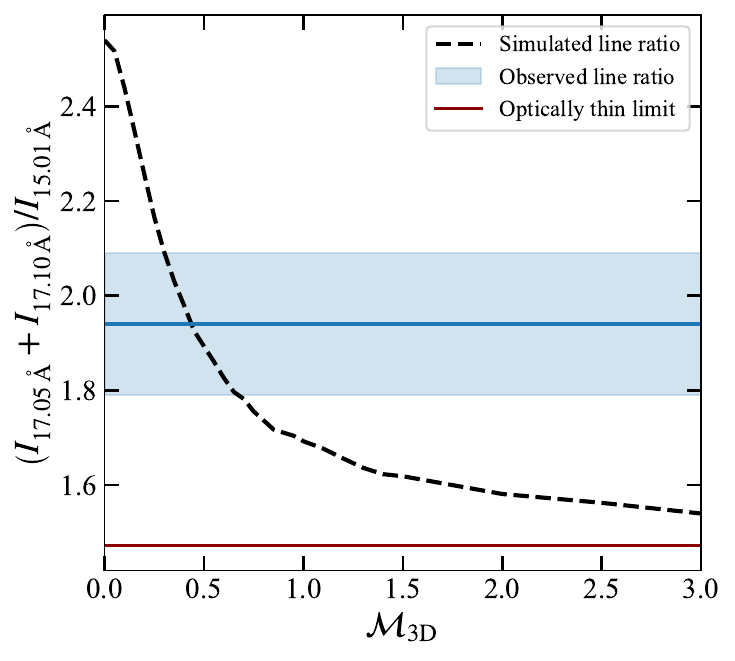}  \includegraphics[width=0.495\linewidth]{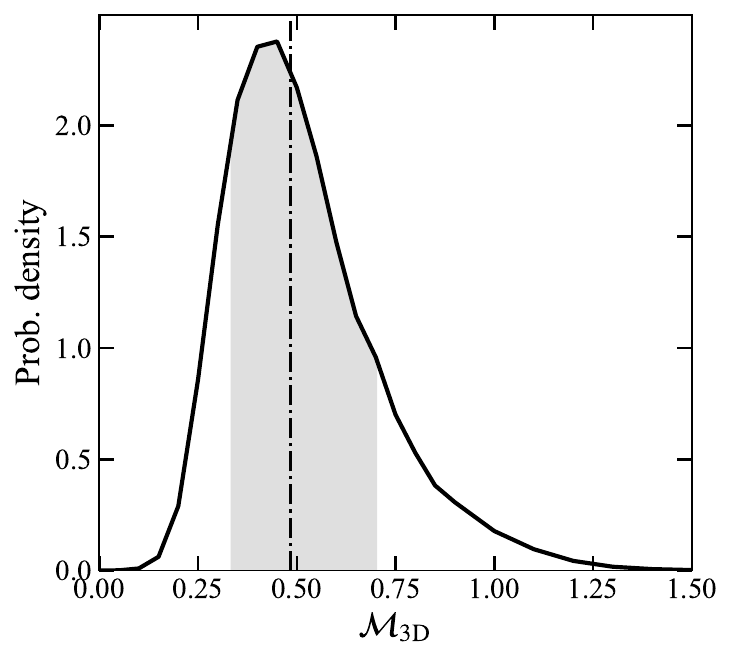}
        \caption{{\bf Left:} The predicted flux ratio $(I_{17.05,\text{\AA}}+I_{17.10,\text{\AA}})/I_{15.01,\text{\AA}}$ as a function of the three-dimensional Mach number. The black dashed curve denotes the simulated flux ratio obtained from the Monte Carlo resonant scattering analysis. The blue line and shaded region show the observed line ratio and its uncertainty measured with RGS. The horizontal maroon line marks the value of the flux ratio in the optically thin limit, which would be approached at high turbulence levels where resonant suppression of Fe~{\sc xvii} 15.01~\text{\AA} is minimal. {\bf Right:} Probability density function (black curve) of the three-dimensional Mach number of the hot gas, by multiplying the independent PDFs from the Fe~{\sc xvii} resonant scattering and \emph{XMM}/RGS line broadening results. The grey shaded region denotes the central 68$\%$ probability interval. The vertical dot-dashed line shows the median $\mathcal{M}_{\rm 3D}=0.48^{+0.22}_{-0.14}$.}
          \label{fig:ifr}
\end{figure*}

For each line, $10^5$ photons are generated from the emissivity-weighted 3D volume, launched in random directions. We assign an initial unit weight to each photon. 
The weight is reduced after each scatter by a factor of $1-e^{-\tau}$. The photon, now carrying its reduced weight, is then scattered into a new direction drawn from a Rayleigh phase function. 
The photons continue to propagate and scatter until their weight falls below $10^{-9}$. 
The final weights of the $17.05~\AA$ and $17.10~\AA$ photons are scaled by their emissivity relative to the $15.01~\AA$ line in the optically thin limit. For a grid of assumed 3D Mach numbers, we compute the total weighted flux of escaped photons within a rectangular aperture matching the RGS extraction region, resulting in the expected line ratio, $(I_{17.05\ \rm \AA}+I_{17.10\ \rm \AA})/I_{15.01\ \rm \AA}$, as a function of the three-dimensional Mach number, as shown in Fig.~\ref{fig:ifr}-left. 
The line ratio approaches the optically thin limit at the high-turbulence end as expected. 

\begin{table*}[h!]
\caption{Observation log of the archived observations targeting NGC 1404 by VLT/MUSE, {\it Chandra} and {\it XMM-Newton} presented in this work. The effective exposure time listed for RGS is an average of the cleaned exposure time of RGS1 and RGS2.}
\label{tab: xmm-record}
\centering
\begin{tabular}{cccc}
\hline
Instrument & ObsID & Date & effective exposure (ks)\\
\hline
{VLT/MUSE} & 296.B-5054(A) & 2017-11-22 &  720$~\times10^{-3}$~($\times5$) \\
\hline
{\it Chandra}/ACIS-S  & 2942  & 2003-02-13  &  29.2 \\
{\it Chandra}/ACIS-I  & 4174 &  2003-05-28 &    45.7\\
{\it Chandra}/ACIS-S  & 9798 & 2007-12-24  &    18.3\\
{\it Chandra}/ACIS-S & 9799  & 2007-12-27 &    21.3\\
{\it Chandra}/ACIS-S  & 16231  & 2014-10-20   & 60.5 \\
{\it Chandra}/ACIS-S  & 17541 & 2014-10-23    &  24.7\\
{\it Chandra}/ACIS-S  & 17540 & 2014-11-02   & 28.5 \\
{\it Chandra}/ACIS-S  & 16233  & 2014-11-09    & 98.8 \\
{\it Chandra}/ACIS-S  & 17548  & 2014-11-11    & 48.2 \\
{\it Chandra}/ACIS-S  & 16232 & 2014-11-12    &  69.1\\
{\it Chandra}/ACIS-S  & 17549  &  2015-03-28   &61.7  \\
\hline
{\it XMM-Newton}/RGS & 0304940101 & 2005-07-30 &  21.3 \\
{\it XMM-Newton}/RGS & 0781350101 & 2016-12-29 &  119.1 \\
\hline
\end{tabular}
\end{table*}

\begin{table*}[h!]
\caption{Best-fit RGS parameters for ObsID 0781350101 extracted from the 90\% PSF region. The \texttt{lpro} scale factor is fixed at $s=1$. Here $F_{17.08\,\AA}$ denotes the combined flux of the unresolved Fe~{\sc xvii} 17.05~\AA\ and 17.10~\AA\ blend.}
\label{tab: rgs fitting result-PSF90}
\centering
\renewcommand{\arraystretch}{1.25}
\setlength{\tabcolsep}{5pt}

\begin{minipage}[t]{0.485\textwidth}
\vspace{0pt}
\centering

\begin{tabular*}{\linewidth}{@{\extracolsep{\fill}}l|lc}
\hline\hline
\multicolumn{3}{c}{Thermal-model parameters} \\
\hline
mod & parameter & value \\
\hline

\multirow{9}{*}{\texttt{cie}}
& norm ($10^{70}\ \mathrm{m^{-3}}$)
& $1.52^{+0.07}_{-0.07}$ \\

& $T$ (keV)
& $0.69^{+0.01}_{-0.01}$ \\

& $v_{\rm rms}$ ($\mathrm{km\,s^{-1}}$)
& $<177$ \\[0.4em]

& \multicolumn{2}{c}{} \\[0.4em]

& N abundance
& $1.57^{+0.36}_{-0.33}$ \\

& O abundance
& $0.49^{+0.05}_{-0.04}$ \\

& Ne abundance
& $0.88^{+0.09}_{-0.09}$ \\

& Mg abundance
& $0.45^{+0.07}_{-0.07}$ \\

& Fe abundance
& $0.42^{+0.02}_{-0.02}$ \\

\hline
\end{tabular*}
\end{minipage}
\hfill
\begin{minipage}[t]{0.485\textwidth}
\vspace{0pt}
\centering

\begin{tabular*}{\linewidth}{@{\extracolsep{\fill}}l|lc}
\hline\hline
\multicolumn{3}{c}{Line measurements and fit statistics} \\
\hline
mod & parameter & value \\
\hline

\multirow{4}{*}{
    \shortstack{
        \texttt{delt}\\
        ($10^{48}\ \mathrm{ph\,s^{-1}}$)
    }
}
& 15.01~\AA\ norm
& $5.52^{+0.37}_{-0.36}$ \\

& 15.26~\AA\ norm
& $2.59^{+0.35}_{-0.33}$ \\

& 16.78~\AA\ norm
& $5.89^{+0.34}_{-0.33}$ \\

& 17.08~\AA\ norm
& $12.21^{+0.47}_{-0.46}$ \\[0.4em]

\cline{1-3}

Flux ratio
& $F_{17.08~{\rm \AA}}/F_{15.01~{\rm \AA}}$
& $1.94^{+0.15}_{-0.15}$ \\[0.4em]

\multirow{4}{*}{Fit}
& C-stat
& 580 \\

& Expected C-stat
& $383\pm28$ \\

& AIC
& 612 \\

& $\Delta\mathrm{AIC}$
& 3 \\

\hline
\end{tabular*}
\end{minipage}
\end{table*}

We combine our resonant scattering and line broadening constraints by first converting the $s$-frozen line broadening $C$-statistic grid from observation 0781350101 into a relative likelihood 
\begin{equation}
p_{\rm LB}(\mathcal{M}_{\rm 3D})\propto
\exp\left[-\frac{\Delta C(\mathcal{M}_{\rm 3D})}{2}\right]
\end{equation}
where $\Delta C=C-C_{\rm min}$. The sampled velocities are converted to three dimensional Mach numbers using $\mathcal{M}_{\rm 3D}=\sqrt{3}v_{\rm rms}/c_s$. The resonant scattering likelihood, $p_{\rm RS}(\mathcal{M}_{\rm 3D})$, is constructed by comparing the measured Fe {\sc xvii} ratio with its predicted value at each $\mathcal{M}_{\rm 3D}$. After interpolating the line broadening likelihood onto the resonant scattering $\mathcal{M}_{\rm 3D}$ grid, we calculate the joint posterior as
\begin{equation}
p_{\rm joint}(\mathcal{M}_{\rm 3D})\propto
p_{\rm LB}(\mathcal{M}_{\rm 3D})
p_{\rm RS}(\mathcal{M}_{\rm 3D})
\end{equation}
We assume a uniform prior over $\mathcal{M}_{\rm 3D}\in[0,3]$, assuming the two RGS constraints are conditionally independent. Our joint posterior, shown in Fig~\ref{fig:ifr}-right, gives $\mathcal{M}_{\rm 3D}=0.48^{+0.22}_{-0.14}$ corresponding to a 1D turbulent velocity $v_{\rm RGS,1D}=120^{+54}_{-38}\ {\rm km\,s^{-1}}$. The central values are posterior medians and the uncertainties capture the central 68\% probability intervals.


\section{Discussion} 
\label{sec:discussion}

\subsection{Origin of off-center warm ionized gas}

In the following subsection, we infer the origin of the off-center warm gas of NGC~1404. Determining whether warm gas has been acquired internally (via in-situ cooling of the hot phase) or externally (via accretion or mergers) can provide insights into the system's assembly history, ISM microphysics, and the overall baryon cycle.


\begin{table*}[t]
\caption{Comparison of the physical properties between the warm ionized and hot gas.}

\label{tab:phase_properties}
\centering
\renewcommand{\arraystretch}{1.35}
\setlength{\tabcolsep}{10pt}

\begin{tabular}{lccc}
\hline\hline
Property
& Warm ionized gas
& Hot gas (overlapping warm gas)
& Hot gas (core) \\
\hline
$T$ (K)
& $10^{4}$ 
& \multicolumn{2}{c}{$8.1\times10^{6}$ (0.70 keV)}
\\

$n_e$ ($\mathrm{cm^{-3}}$)
& $842^{+1046}_{-371}$
& $\approx0.06$  
& $\approx0.65$  \\

$P_{\rm thermal}$ ($10^{-10}\ \mathrm{erg\,cm^{-3}}$)
& $23^{+29}_{-10}$
& $\approx1.3$
& $\approx14$\\

$12+\log(\mathrm{O/H})$
& $8.79^{+0.02}_{-0.02}$
& $\approx8.50$--$8.79$ 
& $\approx8.60$--$8.87$ \\

Velocity ($\mathrm{km\,s^{-1}}$)
& 72
  
& \multicolumn{2}{c}{$120^{+54}_{-38}$}
   \\

\hline
\end{tabular}

\vspace{0.15cm}
\begin{minipage}{0.96\textwidth}
\footnotesize
\textit{Notes.}
The warm gas electron density interval represents the 16th-84th percentile distribution among the 151 spaxels with secure [S\,{\sc ii}] density
measurements. We adopt the LINER calibration to determine the warm gas oxygen abundance. The velocity value for the warm gas is the largest sampled value in its MUSE VSF. 
We assume an isothermal hot atmosphere and the oxygen abundances are inferred using the radially resolved Fe profiles from \cite{2022MNRAS.511.3159M}. We adopt their central and second radial bin for the core and warm gas overlap region, respectively. The range for the hot gas abundances account for both statistical uncertainties in the iron abundance and systematic differences between the oxygen-to-iron ratio found in \cite{2022MNRAS.511.3159M} and ours. For both the hot gas in the warm gas region and that in the core, we list the 1D turbulent velocity corresponding to the median Mach number inferred from our joint resonant scattering-line broadening posterior.

\end{minipage}
\end{table*}

\subsubsection{Radiative cooling and condensation}

The rotation axis of the warm nebula is strongly misaligned with that of the stellar component (Fig.~\ref{fig:continnum}). Gas supplied by stellar mass loss would be expected to inherit the angular momentum of the stars, so this misalignment argues against a stellar origin for the warm gas. A new magnetohydrodynamic (MHD) simulation of NGC~1404's infall into the Fornax cluster (Marin-Gilabert et al., in preparation) shows that the initially non-rotating hot ISM develops rotation as a result of the galaxy's orbital motion, with a rotation axis approximately aligned with that of the observed warm gas. Moreover, the observed VSF of the warm gas approaches the velocity dispersion of the hot gas at large separations, and both lie below the stellar velocity dispersion (Fig.~\ref{fig:vel}). This suggests that the warm gas is kinematically coupled to the hot atmosphere rather than to the stars. 

The off-center warm gas in NGC~1404 may have formed through radiative cooling and condensation of the hot ISM. One mechanism by which perturbations in a hot atmosphere can develop into a multiphase medium is thermal instability \citep{field1965thermal,balbus1986local}. In a gravitationally stratified galactic atmosphere, buoyancy can oppose the growth of condensates by providing a restoring force on displaced gas. The stabilizing role of buoyancy arises as follows: in a convectively stable atmosphere, a gas parcel displaced adiabatically outward retains its original, lower entropy; at its new location it is therefore denser than the ambient medium at the same pressure and experiences a restoring buoyant force. This response, characterized by the Brunt--V\"{a}is\"{a}l\"{a} frequency of buoyant oscillations, is set by the radial entropy gradient and the gravitational acceleration.

Numerical work by \citet{sharma2012thermal} and \citet{mccourt2012thermal} showed that the competition between radiative cooling and the gravitational response of the atmosphere plays an important role in determining whether thermal instability produces a multiphase medium. Although not identical to the buoyancy timescale, the free-fall time, $t_\mathrm{ff}$, is commonly adopted as a dynamical proxy because it captures the gravitational response of the atmosphere. The ratio of the cooling time to the free-fall time, $t_\mathrm{cool}/t_\mathrm{ff}$, is widely used to assess the feasibility of condensation. 
Extended multiphase gas is generally found in cluster atmospheres with $t_\mathrm{cool}/t_\mathrm{ff}\lesssim10$ \citep{Olivares_2019}. However, this threshold is not strict and depends on other conditions such as the entropy stratification of the atmosphere and the amplitude of density perturbations \citep{voit2017global}.

In AGN-driven uplift models, low-entropy gas from the core of the atmosphere is lifted to larger radii in the wakes of rising radio bubbles, where it can become unstable and condense before buoyancy restores it to its equilibrium position \citep{li2014modeling,mcnamara2016mechanism,voit2017global}. 
The simulation tailored to NGC~1404 shows that, rather than moving as a single coherent body, the hot ISM exhibits spatial variations in velocity between neighboring regions, even at the galaxy center (Marin-Gilabert et al., in preparation). Gas parcels therefore move relative to one another as well as relative to the stars and dark matter, allowing low-entropy gas at the center to be displaced from the core, analogous to AGN-driven uplift. Although the simulation does not include radiative cooling, we explore a scenario in which low-entropy gas from the core is ``uplifted'' by RPS and later condenses off-center in NGC~1404.

We calculate the radiative cooling time of the hot ISM as
\begin{equation}
t_\mathrm{cool} = \frac{3 (n_e+n_i)k_\mathrm{B}T}{2 n_e n_\mathrm{H}\Lambda(T,Z)},
\end{equation}
where $n_e$, $n_i$, and $n_\mathrm{H}$ are the electron, ion, and hydrogen number densities, respectively, and $\Lambda(T,Z)$ is the cooling function. Using an APEC plasma model with $k_\mathrm{B}T=0.7$\,keV and $Z=0.4\,Z_\odot$, we obtain $\Lambda(T,Z)=1.88\times10^{-23}\,\mathrm{erg\,s^{-1}\,cm^{3}}$. We approximate the local free-fall time for an isothermal gravitational potential as
\begin{equation}
t_\mathrm{ff} \simeq \frac{r}{\sigma_*},
\end{equation}
where $\sigma_*$ is the stellar velocity dispersion, for which we adopt $\sigma_*\simeq270\,\mathrm{km\,s^{-1}}$. We evaluate $t_\mathrm{ff}$ at the radius of the warm nebula throughout, which gives $t_\mathrm{ff}\simeq5.4$\,Myr. 
For the central hot gas, with $n_e\simeq0.65\,\mathrm{cm^{-3}}$, we obtain $t_\mathrm{cool}\simeq10$\,Myr and $t_\mathrm{cool}/t_\mathrm{ff}\simeq2$. For the hot gas at the radius of the nebula, with $n_e\simeq0.06\,\mathrm{cm^{-3}}$, we obtain $t_\mathrm{cool}\simeq108$\,Myr and $t_\mathrm{cool}/t_\mathrm{ff}\simeq20$. As the uplifted gas expands adiabatically or mixes with the ambient gas, its cooling time may increase. The central values thus provide lower limits and the ambient values at the location of the warm nebula provide upper limits, giving $t_\mathrm{cool}\sim10$--$108$\,Myr and $2\lesssim t_\mathrm{cool}/t_\mathrm{ff}\lesssim20$. This supports the possibility that RPS-driven uplift of low-entropy core gas produces an offset condensate. 

We further estimate the time needed for the low-entropy gas to travel to the position of the warm nebula. The simulation tailored to NGC~1404 predicts that gas at the center is displaced at $v_\mathrm{up}\sim20-50\,\mathrm{km\,s^{-1}}$ (Marin-Gilabert et al., in preparation). Given the projected distance of the warm nebula from the center (1.5\,kpc) and the probable inclination angle of 33$^{\circ}$ relative to the plane of the sky \citep{Su2017}, we estimate a travel time of $t_\mathrm{travel}\simeq35-85$\,Myr. This lies within the range of cooling times, so  
the uplifted gas can plausibly cool and condense en route to its present location.

We also consider a scenario in which turbulent motions generate finite-amplitude density perturbations that cool and condense locally, at the nebula's present location \citep{gaspari2018shaken}. \citet{voit2018role} showed that turbulence can convert buoyant oscillations into condensation in systems where the turbulent velocity dispersion is roughly half the stellar velocity dispersion of the galaxy. For NGC~1404, the one-dimensional velocity dispersion of the hot gas, $120^{+54}_{-38}$\,km\,s$^{-1}$, is indeed approximately half the stellar velocity dispersion, $\sigma_*\simeq270$\,km\,s$^{-1}$. However, at the position of the warm nebula, $t_\mathrm{cool}/t_\mathrm{ff}\sim20$. Using idealized hydrodynamic simulations, \citet{wibking2025a} found that turbulence can trigger condensation only in atmospheres with $t_\mathrm{cool}/t_\mathrm{ff}\lesssim5$, 
a condition not quite met at the current position of the nebula. 
Magnetic fields may alleviate this tension. 
Using numerical simulations of magnetized, stratified atmospheres, \citet{wibking2025b} found that cold gas can form in atmospheres with $t_\mathrm{cool}/t_\mathrm{ff}\simeq20$ only for very strong magnetic fields with the plasma $\beta$ (the ratio of thermal to magnetic pressure) $\lesssim10$. 
Furthermore, as shown in Table~\ref{tab:phase_properties}, the warm gas metallicity agrees better with that of the hot gas at the galaxy center than with that of the hot gas at the same radius as the warm gas. We therefore favor the scenario in which low-entropy hot gas originally at the center of the system was displaced to larger radii by RPS and subsequently cooled and condensed. We note that the warm gas is overpressured relative to its surroundings. This may indicate that it was formed recently and has not yet reached pressure equilibrium.

\subsubsection{Accretion or merger-related acquisition}

External interactions have long been invoked as a key mechanism for building the warm gas reservoir in ETGs \citep{davis2011atlas3d}. 
Here, we explore the feasibility of NGC~1404's offset warm ionized gas having been acquired in an interaction via (1) an infalling satellite system, or (2) a more substantial galaxy merger. 

A well-established correlation exists between stellar mass and interstellar metallicity, with more massive galaxies exhibiting higher oxygen abundances \citep{tremonti2004origin}. Low-mass systems, such as dwarf galaxies, are more susceptible to metal loss due to their shallow gravitational potentials. If the warm ionized gas in NGC~1404 originates from the infall of a dwarf galaxy, we would therefore expect it to be metal-poor relative to the ISM of a massive galaxy. As listed in Table~\ref{tab:phase_properties}, we obtain a high metallicity of 8.79 for the warm gas. In contrast, a predominantly metal-poor, low-mass  (log ($M_*$/M$_\odot$)\,$< 9$) donor is expected to have a metallicity of $<8.6$ from the mass-metallicity relation of \citep{tremonti2004origin}. 
However, the relation of \citet{tremonti2004origin} was established for star-forming galaxies, where star formation is the primary ionization source, and it is unclear whether it is applicable to LINER-like ionization. Furthermore, LINER-like ionization is not what we typically expect for a dwarf galaxy. Overall, we find no compelling evidence that the warm gas originates from the infall of a dwarf galaxy. 



An intermediate-to-major merger with a more massive, chemically enriched galaxy may have delivered the seemingly metal-rich gas \citep{davis2011atlas3d}. A separate line of evidence for this may be found in NGC~1404's stellar kinematic structure. As shown in panel (c) of Fig.~\ref{fig:continnum}, NGC~1404's stellar velocity field shows ordered rotation, with a counter-rotating nuclear core. Furthermore, its stellar velocity dispersion (Fig.~\ref{fig:continnum}; panel (d)) displays bimodal peaks along its major axis. In the ATLAS$^{3D}$ survey \citep{cappellari2011atlas3d, krajnovic2011atlas3d}, a small subset of early-type galaxies have been found to exhibit this counter-rotating velocity field and two-peak velocity dispersion fields. One interpretation is that such systems may be composed of two comparable mass stellar disks with opposite angular momenta (also see \citealt{rix1992ngc}). Galaxy merger simulations have reproduced the observed bimodal velocity dispersion peaks through 1-3:1 mergers \citep{bois2011atlas3d}. It is thus possible that NGC~1404's assembly history involved such a merger, acquiring its offset warm gas in the process.

\subsection{Physical implications of the multiphase gas kinematics}
\label{sec:kinematics}

Regardless of the origin of the warm ionized gas in NGC~1404, it provides an opportunity to probe the kinematics of the ISM. Its VSF, reaching the global scale, complements the hot gas kinematics obtained with \textit{Chandra} and \textit{XMM-Newton} RGS measurements. This multiwavelength diagnosis can place unique constraints on gas kinematics across different gas phases and various spatial scales within a galaxy undergoing environment-driven perturbations.

The warm gas VSF appears to indicate two possible driving scales. The first is at 
200 pc, which is the scale expected for perturbations driven by bubbles from Type Ia supernovae (SNIa). This is consistent with NGC~1404 being an early-type galaxy, in which the supernova population is dominated by SNIa. The other is possibly beyond the largest sampled spatial scales at $\gtrsim$1\,kpc, which is presumably related to the ongoing infall and stripping of NGC~1404.

RGS measures a hot-gas, one-dimensional velocity dispersion of $120^{+54}_{-38}$\,km/s. This measurement agrees well with the peak velocity derived from the power spectrum of the {\it Chandra} surface-brightness fluctuations for the same slab as the RGS analysis (Fig.~\ref{fig:vel}, dark blue band). 
We further used {\it Chandra} to derive the velocity for a much smaller region which overlaps with the extent of the warm gas. As shown in Fig.~\ref{fig:vel}, for the same region, while the hot gas kinematics overall exceeds that of the warm gas, the velocities of the two phases are consistent at 0.2 and 0.7\,kpc, indicating some degree of kinematic coupling between the two phases. 
Also based on the \emph{Chandra} measurements, at the same radii from the center of the galaxy, the hot gas velocity amplitudes in the downstream regions are larger than those measured in other directions, which could be envisioned as enhanced gas perturbations induced by the infall and RPS of NGC~1404. 


The infall of NGC~1404 through the Fornax cluster's ICM is a plausible driver of its gas motion. Hydrodynamical simulations tailored to the interaction between NGC~1404 and Fornax suggest that a projected wake lies to the south of NGC~1404, and should show enhanced turbulence \citep{sheardown2018recent}. Furthermore, the structure of this interaction has been probed using deep \emph{Chandra} observations, revealing sub-kpc Kelvin-Helmholtz Instabilities (KHIs) at the leading northern edge as well as mixing in the downstream tail between NGC~1404's cold ISM and the hot ICM \citep{Su_etal_2017}. Simulations of stripped ellipticals have shown that KHI structures along the sides of the remaining hot ISM can contribute to gas motions on kpc scales \citep{roediger2015stripped}. From Fig.~15 in \cite{sheardown2018recent}, we can estimate a transverse radius of $\sim5-7$ kpc for NGC~1404's remaining hot ISM. Assuming (for illustration) the KHIs grow to approximately 1/3 of this radius, the eddies would reach to $\sim1.7-2.3$ kpc, within a factor of $\sim2$ to the potential VSF turnover location at $\sim1$ kpc (Fig.~\ref{fig:vel}). 
This may suggest that such shear-driven disturbances can influence gas motions on kpc scales. 


Our kinematic measurements allow us to infer the turbulent kinetic and thermal energy budget of NGC~1404's hot atmosphere. Assuming that turbulence is isotropic, the ratio of turbulent to thermal energy density is

\begin{equation}
    \frac{\epsilon_{\mathrm{turb}}}{\epsilon_{\mathrm{th}}}=\gamma\left(\frac{\mathcal{M}_{\rm 3D}}{\sqrt{3}}\right )^2 .
\end{equation}
Using our estimated $\mathcal{M}_{\rm 3D}=0.48^{+0.22}_{-0.14}$, we find $\epsilon_{\mathrm{turb}}/\epsilon_{\mathrm{th}}=0.13^{+0.14}_{-0.07}$. 
\cite{2017MNRAS.472.1659O} performed a systematic study of a sample of nearby massive elliptical galaxies via resonant-scattering and line-broadening analysis using {\it XMM-Newton} RGS. The authors found a best-fit mean 1D velocity dispersion of $\sim110$\,km\,s$^{-1}$ and $\epsilon_{\mathrm{turb}}/\epsilon_{\mathrm{th}}\sim0.056$. Many of the systems in \cite{2017MNRAS.472.1659O} are cool-core massive elliptical galaxies with X-ray cavities coincident with radio bubbles, indicating that their global-scale gas motions are primarily driven by mechanical AGN feedback. 
This is in contrast to NGC~1404, whose gas motion on global scales is mainly driven by stripping. 
This finding shows that environmental perturbations can drive ISM turbulence no weaker than that driven by AGN mechanical feedback. 

\section{Conclusions}
\label{sec:conclusions}

We present VLT/MUSE, {\it Chandra}/ACIS, and {\it XMM-Newton}/RGS observations of the multiphase ISM of NGC~1404, a massive elliptical galaxy undergoing ram-pressure stripping in the nearby Fornax Cluster. In addition to its substantial X-ray–emitting hot gas, NGC~1404 also hosts an off-center warm nebula in the south. Our findings are summarized below.

   \begin{enumerate}

    \item The warm nebula extends over 1 kpc and is offset by 1.5 kpc to the south of the galaxy center. The warm gas appears to rotate along the east–west direction, misaligned with the stellar component, which rotates along the north–south direction. Our emission line analysis shows that the warm gas has LINER-like excitation. The inferred oxygen abundance of the warm gas is comparable to that of the hot gas at the galaxy center. 

    \item We derive the first-order VSF of the line-of-sight velocity map of the warm gas after removing the underlying rotation. The VSF at small scales displays a slope of $\approx1/3$, as expected for classical Kolmogorov turbulence. We find that the VSF may exhibit two characteristic driving scales. One is at 200 pc, consistent with the expectation from SNIa-driven bubbles. The other, likely at $\gtrsim1$ kpc, may be driven by the stripping process.

    \item We place constraints on the one-dimensional velocity dispersion of the hot gas on galaxy scales, finding $<177$ km s$^{-1}$ from line broadening measured with {\it XMM-Newton}/RGS. Combining the complete likelihoods from our line broadening and resonant scattering analyses gives $\mathcal{M}_{\rm 3D}=0.48^{+0.22}_{-0.14}$, corresponding to a one dimensional turbulent velocity of $120^{+54}_{-38}~\mathrm{km~s^{-1}}$. 
    We find a turbulent-to-thermal energy density ratio of $\epsilon_{\mathrm{turb}}/\epsilon_{\mathrm{th}}=0.13^{+0.14}_{-0.07}$, which is elevated compared to typical AGN-driven turbulence in early-type galaxies.

   \item We performed a surface-brightness analysis of deep {\it Chandra} observations and derived the power spectra of various regions of interest in the ISM of NGC~1404. We obtain a consistent velocity measurement with the RGS measurement for the same slab used in the RGS analysis. 
   The \emph{Chandra} hot-gas velocity amplitude in the region spatially coincident with the warm gas converges with the warm gas VSF, 
   suggesting some degrees of kinematic coupling between the warm and hot phases. 
   
   \item The off-center warm gas in NGC~1404 may have formed through radiative cooling of the hot ISM. Low-entropy hot gas from the galaxy center may have been uplifted by ram pressure stripping, analogous to AGN-driven uplift, and condenses en route. Bracketing the cooling time of the uplifted gas between the central and ambient values gives $t_\mathrm{cool}\sim10$--$108$\,Myr and $2\lesssim t_\mathrm{cool}/t_\mathrm{ff}\lesssim20$, while the estimated travel time to the nebula, $\simeq35-85$\,Myr, lies within this range. 
   Alternatively, if the warm gas originates externally, our analysis favors a merger with a galaxy of comparable mass over an infalling satellite system.

   \end{enumerate}

\section*{Acknowledgements}

R.E. and Y.S. acknowledge NASA award 80NSSC24K1776 and Chandra awards GO5-26073X and GO5-26051X. 
Y.L. acknowledges support from NASA grant 80NSSC22K0668, Chandra X-ray Observatory grant TM3-24005X, NSF grants AST-2514692, AST-2510198, and CAREER award AST-2516092.

\bibliographystyle{apsrev4-1}

\bibliography{ref.bib}


\end{document}